\documentclass[preprint2]{aastex}

\def\vlsr{$V_{\mbox{\scriptsize LSR}}$}
\def\kms{~km~s$^{-1}$}

\def\etal{~et\ al.\ }
\def\h2o{H$_{2}$O}
\def\iras1846{IRAS~18460$-$0151}
\def\j1833{J183307.8$+$011535}
\def\masyr{~mas~yr$^{-1}$}

\slugcomment{To be submitted to the Astrophysical Journal.}

\shorttitle{\h2o and OH masers in \iras1846.}
\shortauthors{H.~Imai \etal}

\begin{document}

\title{THE SPATIOKINEMATICAL STRUCTURE OF \h2o\ AND OH MASERS IN THE ``WATER FOUNTAIN" SOURCE \iras1846}

\author{Hiroshi Imai\altaffilmark{1,2}, Shuji Deguchi\altaffilmark{3}, 
Jun-ichi Nakashima\altaffilmark{4}, Sun Kwok\altaffilmark{4}, 
and Philip J.~Diamond\altaffilmark{5}}

\altaffiltext{1}{Department of Physics and Astronomy, 
Graduate School of Science and Engineering, Kagoshima University, 1-21-35 Korimoto, 
Kagoshima 890-0065, Japan;  
hiroimai@sci.kagoshima-u.ac.jp}

\altaffiltext{2}{International Centre for Radio Astronomy Research, M468, 
The University of Western Australia, 35 Stirling Hwy, Crawley, Western Australia, 6009, Australia} 

\altaffiltext{3}{Nobeyama Radio Observatory, National Astronomical Observatory of Japan, 
Minamimaki, Minamisaku, Nagano 384-1305, Japan; deguchishuji60@gmail.com}

\altaffiltext{4}{Department of Physics, University of Hong Kong, Pokfulam Road, 
Hong Kong, China; junichi@hku.hk, sunkwok@hku.hk}

\altaffiltext{5}{SKA Organisation, Jodrell Bank Observatory, Lower Withington, Macclesfield, 
Cheshire SK11 9DL, UK; diamond@skatelescope.org}

\received{2012 12 19}
\accepted{2013 07 03}
%\journalid{???}{?? ???? 2013}
%\articleid{??}{??}

\begin{abstract}
Using the Very Long Baseline Array and the European VLBI Network, we have observed 22.2~GHz \h2o\ and 1612~MHz OH masers in the {\it water fountain} source \iras1846. The \h2o maser spectrum has a very wide line-of-sight velocity range ($ \approx$310\kms) and consists of three groups of emission features at the blue-shifted ($-68$\kms$\lesssim$ \vlsr $\lesssim-17$\kms) and red-shifted (\vlsr$\simeq$240\kms) edges as well as around the systemic velocity (112\kms$\lesssim$ \vlsr $\lesssim$133\kms). The first two \h2o\ spectral components exhibit a highly-collimated high-velocity bipolar jet on the sky, with an angular separation of $\approx$120 milliarcseconds (mas) (240~AU in linear length) and a three-dimensional flow velocity of $\approx$160\kms. The flow dynamical age is estimated to be only $\approx$6~yr (at the time of the observation epochs of 2006--2007). Interestingly, the systemic velocity component clearly exhibits a spherically-expanding outflow with a radius of $\approx$36~AU and a flow velocity of $\approx$9\kms. On the other hand, the OH maser spectrum shows double peaks with a velocity separation of $\approx$25\kms (\vlsr$=$111--116 and 138--141\kms), as typically seen in circumstellar envelopes of OH/IR stars. The angular offset between the velocity-integrated brightness peaks of the two high-velocity \h2o components is $\approx$25~mas (50~AU). The offset direction and the alignment of the red-shifted maser spots are roughly perpendicular to the axis of the \h2o\ maser flow. High-accuracy astrometry for the \h2o and OH masers demonstrates that the collimated fast jet and the slowly expanding outflow originate from a single or multiple sources which are located within 15~mas (30~AU). On the other hand, the estimated systemic velocity of the collimated jet ($V_{\rm sys}\approx$87--113\kms) has a large uncertainty. This makes it difficult to provide strong constraints on models of the central stellar system of \iras1846.
\end{abstract}

\keywords{masers---stars: AGB and post-AGB---stars:  individual(\iras1846)--- stars: kinematics---stars: mass loss---stars: winds, outflows}

\section{INTRODUCTION}
``Water fountain'' sources (WFs) are asymptotic giant branch (AGB) or post-AGB stars that host highly-collimated, fast ($V_{\rm exp}\gtrsim$50\kms), bipolar jets traced by 22.2~GHz \h2o\ maser emission. Interestingly, all of the observed dynamical ages of the WF jets are less than 100~yr. As of 2013, 15 sources have been identified as possible WFs (in more detail, see \citealt{ima07a}; \citealt{des12}). It is expected that they are signposts of the transition phase from the AGB to post-AGB, when one can see ignition of a stellar jet and a resulting metamorphosis from spherically symmetric to asymmetric structure of the circumstellar envelope (CSE). 

Here we focus our interest on some WFs, in which \h2o\ masers near the dynamical centers of the jets show lower expansion velocities, roughly equal to a typical expansion velocity of a CSE in the AGB phase ($V_{\rm exp}\lesssim$30\kms), and exhibit spherical symmetric distributions. The spherical expansion in such an expansion velocity is well traced in AGB CSEs by 1612~MHz OH maser emission (see, e.g., the stellar OH maser data base, \citealt{eng12}).\footnote{http://www.hs.uni-hamburg.de/maserdb} They are a completely different component from the collimated WF jet whose velocity often exceeds 100\kms\ (e.g., \citealt{ima02a,ima07a}). For these interesting WFs, it is considered that they still host relic CSEs made at the end of the AGB phase because the majority of WFs are strongly obscured in the near-infrared bands (e.g. \citealt{deg07}). On the other hand, it is also considered that the stellar mass loss rate in the AGB phase will increase over $\dot{M}\geq 10^{-7}\: M_{\sun}{\rm yr}^{-1}$  or much higher \citep{bow94}. A WF likely appears when the mass loss rate of the AGB CSE reaches its maximum, but what will happen to the CSE soon after this phase?  Thus, the exploration of the inner part of CSEs, including the OH and low-velocity \h2o\ masers, is important for identifying the stellar evolutionary phase through diagnosing the kinematics of such stellar mass loss. Eventually, the study of the WFs will provide important clues to understanding the mechanisms of stellar jet launching and planetary nebula shaping in terms of the AGB to post-AGB transition. 

In this paper, we present the spatiokinematics of \h2o\ masers and their spatial relationship with the 1612~MHz OH maser emission associated with \iras1846 (hereafter abbreviated as I18460), as revealed by observations with the Very Long Baseline Array (VLBA) and the European VLBI Network (EVN). The preliminary results were reported by \citet{ima07a} and \citet{ima08}. Here we describe the results in more detail. This paper also presents additional data of the I18460 \h2o\ masers that were used for determining the secular proper motion of I18460. I18460 was first identified as a WF by \citet{deg07}. The \h2o\ maser spectrum showed a very large velocity separation of $ \approx$290\kms, centered at the local-standard-of-rest (LSR) velocity of \vlsr$\approx$100\kms. It also showed the systemic velocity component, whose velocity (\vlsr$\approx$120\kms) coincides with that of the intrinsic CO emission of I18460 \citep{riz12} and the center velocity of the double-peaked spectrum of OH masers \citep{sev97}. Such a systemic velocity component of \h2o\ maser emission has sometimes been observed in the archetypal WF source, W43A \citep{ima02a,ima07a}. It has been unclear with what that component is associated, candidates are either a {\it relic} AGB CSE as defined above or an ``equatorial'' flow. The latter has an expansion velocity ($v_{\rm exp}\gtrsim 30$\kms) higher than that of the former ($v_{\rm exp}\approx$10--20\kms). The existence of an equatorial flow in a WF was recognized in the spatiokinematical structure of SiO masers in W43A (\citealt{ima05} (see also the case of IRAS~16342$-$3814, \citealt{sah99}) ), but its origin is still obscure because of limited data from such flows. In IRAS~18286$-$0959, Paper II also recognized the existence of a similar flow, but \h2o\ maser features associated with the flow were considered as ``outliers" of the collimated jet, which show proper motions much slower than those in the jet. In the case of I18460, we revealed the spatiokinematics of \h2o\ masers in the systemic velocity components in detail for the first time in WFs. In this paper, we adopt a distance to I18460 of $\sim$2~kpc (see Section \ref{sec:center}), which is considerably less than the kinematic distance derived by assuming a circular orbit in the Milky Way (6.0 or 7.8~kpc).  

Section \ref{sec:observations} describes the VLBA and EVN observations of the \h2o\ and  OH masers in I18460, respectively, including normal maser source mapping and phase-referencing astrometric procedures. Section \ref{sec:results} presents the \h2o and OH maser maps and the results of astrometry. Section \ref{sec:discussion} discusses the slowly expanding flow of I18460, whose spatiokinematical structure is clearly elucidated, and the secular motion of I18460 in the Milky Way, which may provide an important clue to revealing the character of the central star in I18460.   

\section{OBSERVATIONS AND DATA REDUCTION}
\label{sec:observations}

We describe the VLBA and EVN observations (project codes BI033 and EI009, respectively) and the analyses of their data, which follow the same procedure as those in Paper II. We also describe the astrometry of the \h2o masers conducted in other VLBA observations (BI036). 

\subsection{VLBA Observations of the \h2o\ Masers}
\label{sec:VLBA}

Table \ref{tab:status} summarizes the status of the VLBA observations and data reduction of the I18460 \h2o\ masers. VLBA observations were conducted three times during 2006 September--2007 February (BI033) and three times during 2008 September--2009 February (BI036). In the first set of observations, another \h2o\ maser source IRAS~18286$-$0959 was also observed at the same time \cite{ima13b} (hereafter Paper II), therefore the on-source time for I18460 was limited to 3~hr. In the second set of observations, the antenna-nodding, phase-referencing technique was employed for astrometry of the I18460 \h2o\ masers, in which the maser and the phase-reference source J185146.7$+$003532 (hereafter abbreviated as J1851, 2$^{\circ}$.52 away from I18460) were observed in a source-switching cycle of one minute. In all of the six sessions, the continuum source OT~081 was also observed for 4 minutes in every 20 minutes for calibration of group-delay residuals and bandpass characteristics. The received signals were recorded at a rate of 128~Mbits~s$^{-1}$ with 2 bit quantization into eight base-band channels (BBCs) in dual circular polarization. The centers of four pairs of BBCs, each of which has a bandwidth of 4~MHz, were set to LSR velocities of $-$55, 0, 120, and 240\kms. The recorded data were correlated with the Socorro FX correlator using an integration period of 2~s. The data of each BBC was divided into 256 spectral channels, yielding a velocity spacing of 0.21\kms\ per spectral channel. The following coordinates were adopted as the delay-tracking centers in the data correlation:

$\alpha_{\rm J2000}=$18$^{\rm h}$48$^{\rm m}$42$^{\rm s}$\hspace{-2pt}.94,
$\delta_{\rm J2000}=$~$-$01$^{\circ}$48$^{\prime}$30\farcs 28 and 

$\alpha_{\rm J2000}=$18$^{\rm h}$48$^{\rm m}$43$^{\rm s}$\hspace{-2pt}.027,
$\delta_{\rm J2000}=$~$-$01$^{\circ}$48$^{\prime}$30\farcs 46 for I18460 for the first and second sets of three sessions, respectively, and
 
$\alpha_{\rm J2000}=$18$^{\rm h}$51$^{\rm m}$46$^{\rm s}$\hspace{-2pt}.723080, 
$\delta_{\rm J2000}=$~$+$00$^{\circ}$35$^{\prime}$32\farcs 36283 for J1851. 

For the data reduction, we used the NRAO's AIPS package and employed a normal procedure. After calibration of group-delay residuals and bandpass characteristics using scans on OT~081, fringe fitting and self-calibration were made using a spectral channel that contained bright maser emission. Column 3 in Table \ref{tab:status} gives the LSR velocity of the spectral channel selected as phase- and position-reference. The obtained solutions of calibration were applied to the data in all spectral channels. Finally, image cubes of the maser source were synthesized in visibility deconvolution through the CLEAN algorithm. Because most of the \h2o\ maser spots (velocity components) were spatially partially resolved,  calibration solutions and the associated visibilities were missing from relatively long baselines ($B_{\lambda}\gtrsim {\rm 200~M}\lambda$) in the image synthesis. As a result, naturally weighted visibilities associated with available self-calibration solutions yielded a typical synthesized beam of 1.5~mas$\times$1.0~mas. Columns 4 and 5 of Table \ref{tab:status} give values of the 1 $\sigma$ noise level and the synthesized beam pattern, respectively. 

In each spectral channel of the synthesized image cube, using a POPS pipeline script including the AIPS task SAD, brightness peaks higher than 7 $\sigma$ noise level were identified as {\it maser spots} and their parameters as Gaussian brightness components were extracted. We determined the position of a {\it maser feature},  which is a physical maser clump consisting of a cluster of maser spots seen in consecutive velocity components and at almost the same position within the synthesized beam, as described in \citet{ima02b}. The peak-brightness velocity of a feature was determined by quadratic fitting of the peak intensities of the brightest three spots in the feature, then the position of the corresponding peak velocity was determined from linear interpolation of the positions of the two brightest spots. 

For the astrometry of the I18460 \h2o masers, in the sessions of 2008--2009, all phase calibration steps were made using the data of J1851, the solutions were then applied to the data of I18460. Because the calibration of unknown zenith delay residuals was imperfect and I18460 is very close to the celestial equator, the synthesized maser source images were partially defocused, especially in the north--south direction. However, the obtained position accuracy of the \h2o\ masers ($\sim$0.1~mas and $\sim$0.5~mas in the right ascension (R.A.) and declination directions (decl.), respectively) is sufficient to measure the linear proper motions of the masers and to compare their coordinates with those of the  OH masers. For measurement of {\it relative} maser proper motions, we used the image cube obtained from the astrometric analysis rather than that from the self-calibration scheme mentioned above. 

\notetoeditor{Put Table 1 and Figure 1 here.}

\subsection{EVN Observation of the OH Masers}
\label{sec:EVN}

The EVN observation of the I18460 OH masers ($F_{j}=1^{+}\rightarrow 2^{-1}$ $^{2}\Pi_{3/2}$ at 1.612231f~GHz) (project code EI009B) was conducted on 2007 June 11 using the following telescopes: Lovell 76~m in Jodrell Bank, Cambridge 32~m, Westerbork Synthesis Radio Telescope (14 25-m), Medicina 32~m,  Noto 32~m, Effelsberg 100~m, Onsala 25~m, Torun 32~m, and Hartebeesthoek 26~m. We used the antenna-nodding phase-referencing technique (e.g., \citealt{bea95}) to obtain astrometric measurements for the maser source. This was possible for every telescope except the Lovell 76~m, which always observed I18460 without antenna nodding. I18460 and the phase-reference continuum emission source J183307.8$+$013511 (hereafter abbreviated as J1833) were scanned in a source-switching cycle of 5~minutes, resulting in the total duration of scans on I18460 being $\sim$3~hr. J183243.5$+$135744 was observed for 3 minutes in every 40 minutes for calibration of group-delay residuals and bandpass characteristics. The received signals were recorded at a rate of 128~Mbits~s$^{-1}$ with 2 bit quantization into eight BBCs in dual circular polarization. One pair of BBCs had a bandwidth of 1~MHz and others have 4~MHz for observing the I18460 \h2o\ maser and the J1833 continuum emission, respectively. 

The recorded data were correlated with the Mark IV processor in the Joint Institute for VLBI in Europe using an integration period of 4~s. The data of each BBC was divided into 512 and 32 spectral channels, respectively, for the maser and continuum observations. A velocity channel spacing of 0.36\kms\ was obtained for the maser data. The coordinates of I18460 adopted in the data correlation were the same as the first set of the coordinates for the VLBA data (Section \ref{sec:VLBA}). The following coordinates for J1833 were used in the data correlation:

$\alpha_{\rm J2000}=$18$^{\rm h}$33$^{\rm m}$07$^{\rm s}$\hspace{-2pt}.760875,

$\delta_{\rm J2000}=$~$+$01$^{\circ}$15$^{\prime}$35\farcs 30098.  

In the analysis of the EVN data, in addition to the standard calibration procedures as described in Section \ref{sec:VLBA}, calibration of residual delays due to the ionosphere was also applied using the AIPS task TECOR.\footnote{There are several global ionosphere models provided for this correction, but their results were similar (e.g., \citealt{vle07}).} Fringe fitting and self-calibration were carried out as done for our VLBA observations but using the data of J183243.5$+$135744. Phase referencing, fringe fitting and self-calibration were undertaken using the data of J1833. The calibration solutions were applied to the data of I18460. The image of J1833 and the image cube of the OH maser source were synthesized in visibility deconvolution through the CLEAN algorithm. Calibration solutions and their associated visibilities were lost from long baselines ($B_{\lambda}\gtrsim {\rm 10~M}\lambda$) on J1833, which has an extended structure; this resulted in a large beam size for the image synthesis of the OH maser source. Through the phase-referencing technique, we synthesized a maser image cube (referred to as the phase-referencing-based cube) and determined the absolute coordinates of the OH maser source.

In addition, in order to obtain a higher angular resolution image cube of the OH maser source, free from the effects of the extended structure of J1833, fringe fitting and self-calibration were also carried out using the OH maser data. The spectral channel at \vlsr$=$111.7\kms\ was used as phase reference. The uniformly weighted visibilities yielded a synthesized beam of 97~mas$\times$23~mas at a P.A.$=$ 11\arcdeg. Thus we generated another maser image cube (referred to as the self-calibration-based cube). The OH masers were also spatially resolved on baselines to the Hartebeesthoek station, causing the elongated beam pattern in the north--south direction.

\notetoeditor{Put Table 2 and Figure 2 here. }

\section{RESULTS}
\label{sec:results}
Tables \ref{tab:features_h2o_a} and \ref{tab:features_h2o_b} give the parameters of \h2o\ maser features detected in the observation sessions in 2006--2007 and 2008--2009, respectively. In Table \ref{tab:features_h2o_b}, the  positions are given as offsets with respect to the phase-tracking center of I18460.

With LSR velocity and proper motion measurements, the three-dimensional spatiokinematics of the \h2o\ masers has been revealed. Table \ref{tab:pmotions} gives the list of {\it relative} proper motions of \h2o\ maser features found in all observations. Note that the position-reference maser features \iras1846:I2013 {\it 6} and {\it 19} have similar LSR velocities but are unlikely the same maser feature. They were observed at different epochs and likely located around the eastern and southern parts of the center maser region, respectively. It is difficult to find the same feature detected in both of the first three epochs (2006--2007) and the last three epochs (2008--2009). Therefore, it is difficult to combine the proper motion data obtained from the 2006--2007 season with those from the l2008--2009 season. Nevertheless, we hereafter analyze the two sets of proper motions independently while checking their mutual consistency in the maser distributions and the kinematics.

The \h2o\ masers clearly exhibit both a highly collimated bipolar jet in the north--south direction (see Section \ref{sec:jet})  and a central cluster of maser features (hereafter the central components, see Section \ref{sec:center}). The latter masers have LSR velocities within 10\kms\ of the systemic velocity, which we adopt as \vlsr$=$124.7\kms\ \citep{sev97} in this paper. Using the astrometric data obtained during 2008 September--2009 February, the absolute coordinates of the \h2o\ masers in the central components and the secular stellar proper motion were derived (see Section \ref{sec:secular-motion}). Using the same data, an annual parallax of the \h2o\ masers was tentatively detected. 

\begin{figure}
\epsscale{1.0}
\plotone{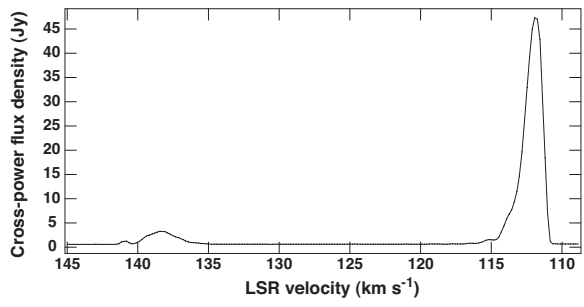}
\caption{Cross-power spectrum of the OH masers in \iras1846 obtained from the Effelsberg--Westerbork baseline.\label{fig:oh-spectrum}}
\end{figure}

The OH masers show a double-peaked spectrum with a velocity separation of $\approx$25\kms (\vlsr$=$111--114 and 136--140\kms\ in the phase-referencing-based cube and \vlsr$=$111--116 and 139--140\kms\ in the self-calibration-based cube), which is typical of that seen in other AGB stars. Figure \ref{fig:oh-spectrum} shows the cross-power spectrum of the OH masers in I18460. Table \ref{tab:features_oh} gives the parameters of the OH maser spots detected using the two types of calibration scheme. The reason for the difference in the velocity ranges in which the maser spots are detected is unclear, but we speculate is due to the difference in the dynamic range caused by different conditions in phase calibration and the difference in the synthesized beam patterns.  They are also exactly associated with the central \h2o\ maser components (see Section \ref{sec:OH}). 

\notetoeditor{Put Table 3 and Figure 3 here. }

\subsection{The Collimated Fast Jet}
\label{sec:jet}

\begin{figure*}
\begin{center}
\epsscale{2.0}
\plotone{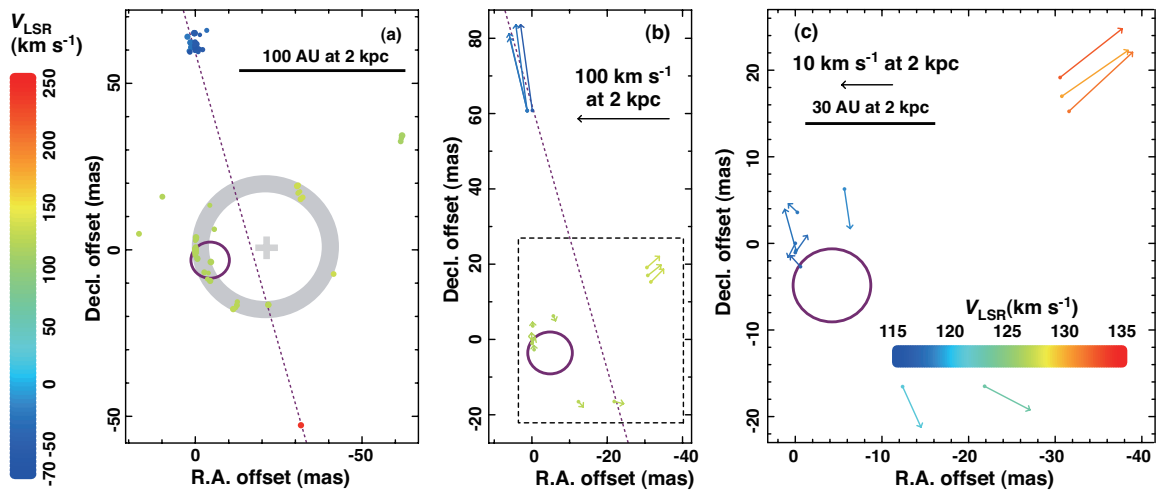}
\end{center}
\vspace*{-5mm}
\caption{\h2o masers found in \iras1846 in 2006--2007. (a) All of the \h2o maser features (dots) detected in all observations. The three-epoch maps are superposed using the common phase-reference maser spot in the maser feature \iras1846:I2013-{\it 6} located at the map origin. A dashed magenta line (shown also in (b)) indicates the axis of the fast jet. A thick gray circle indicates a maser feature ring whose center is marked with a grey plus sign. An opened magenta ellipse (also shown in sub-panels (b) and (c)) indicates the location of the dynamical center of the central spherical flow, which was estimated by model fitting and whose uncertainty is indicated by the size of the ellipse. (b) Relative proper motions of \h2o\ masers (arrows) in the whole region of \iras1846, which were measured with respect to the position-reference maser feature \iras1846:I2013 -{\it 6}. (c) Same as (b) but in the central region shown in a dashed box in (b). The systemic motion ($-$1.4, 2.4)(km~s$^{-1}$) has been subtracted from the originally measured proper motions.    
\label{fig:H2O}}
\end{figure*}

Figure \ref{fig:H2O} shows a comprehensive view of the spatiokinematics of \h2o\ masers in I18460 found in the first three epochs. As shown in Figure \ref{fig:H2O}(a) and (b), the blue-shifted ($-68$\kms$\lesssim$ \vlsr $\lesssim-17$\kms) and red-shifted (\vlsr$\simeq$242\kms) components are separated by $\approx$120~mas, corresponding to a linear scale of $\approx$240~AU at 2~kpc. We detected highly collimated, fast proper motions of the blue-shifted masers ($\mu\approx$10~mas~yr$^{-1}$) with respect to the central masers. However, we could not measure proper motions of the red-shifted masers because they were detected at only the first epoch. Nevertheless we can approximate the relative proper motions of the blue-shifted masers with respect to the central masers as the velocity vectors of one side of the bipolar jet components. The maser locations and motions projected on the sky give a dynamical age of the jet to be only $\approx$6~yr in the observation epochs of 2006 September--2007 February. 

The proper motions are well aligned along the jet axis, which is defined here to be a straight line connecting the blue-shifted masers with the red-shifted. The jet axis passes through the central masers. If we assume a circular distribution of the central masers (a thick grey circle in Figure \ref{fig:H2O}(a) ), the separation between the circle center (a plus symbol in Figure \ref{fig:H2O}(a) ) and the jet axis is only $\approx$10~mas ($\approx$20~AU). Because of the limited number of jet masers, this paper does not discuss a precessing jet model as proposed by \citet{ima05} and \citet{yun11}. 

The center velocity of the two velocity components of the jet masers ranges between $V_{\rm center}\approx$87--113\kms, depending on which pair of the blue-shifted and red-shifted components are linked. Pairing of the \vlsr$\simeq$242 and $-17$\kms\ components gives $V_{\rm center}\approx$113\kms, which is close to the systemic velocities derived from the central components and the OH masers (\vlsr$\sim$125\kms) as mentioned in Sections \ref{sec:center} and \ref{sec:OH}, respectively. However, we cannot rule out pairing the \vlsr$\simeq$242 and $-67$\kms\ components to derive the center velocity. We derive a three-dimensional flow speed of $V_{\rm jet}\approx$180\kms\ at an inclination of $i\approx$32\arcdeg\ with respect to the line of sight under the assumption of $V_{\rm sys}\approx$87\kms\. If we assume $V_{\rm sys}\approx$125\kms , then we derive $V_{\rm jet}\approx$150\kms\ at $i\approx$40\arcdeg\. The implication of the difference in the assumed systemic velocity is discussed in Section \ref{sec:jet2}. In both cases, a position angle of the jet, P.A.$\approx$8\arcdeg\ east from north, is derived. 

\subsection{The Central Slow Flow Traced by \h2o\ Masers}
\label{sec:center}

As briefly mentioned in Section \ref{sec:jet} and as shown in Figure \ref{fig:H2O}(a), the distribution of a large fraction of \h2o\ maser features in the central component is well expressed by a (thick) ring with a radius of $\approx$18~mas ($\approx$36~AU). The maser features located in the central region but outside the ring are distributed in the east--west direction, perpendicular to the jet axis. We are able to measure 11 relative proper motions of \h2o\ masers. As shown in Figure \ref{fig:H2O}(b) and (c), the proper motions are also loosely collimated perpendicular to the jet axis and exhibit an expanding flow. 

In order to derive the kinematical parameters of this central outflow, we undertook a least-squares method based model-fitting analysis as described by \citet{ima00} and \cite{ima12b}. Using the model fitting, we derive a position vector of the originating point of outflow in the maser map, $(\Delta X_{\rm 0}, \Delta Y_{\rm 0})$ and a velocity vector of the originating point, $(V_{0x}, V_{0y})$. They were estimated by minimizing a $\chi^2$ value, 

\begin{eqnarray}
\nonumber
\chi^{2}& = & \frac{1}{3N_{\rm m}-N_{\rm p}}
\sum^{N_{\rm m}}_{i}
\left\{
\frac{\left[\mu_{ix}-w_{ix}/(a_{{\rm 0}}D)\right]^{2}}{\sigma^{2}_{\mu_{ix}}} \right. \\
& & 
\left. +\frac{\left[\mu_{iy}-w_{iy}/(a_{{\rm 0}}D)\right]^{2}}{\sigma^{2}_{\mu_{iy}}}
+\frac{\left[u_{iz}-w_{iz}\right]^{2}}{\sigma^{2}_{u_{iz}}} 
\right\}. 
\label{eq:model-fit}
\end{eqnarray}

\noindent
Here $N_{\rm m}=11$ is the number of maser features with measured proper motions, $N_{\rm p}=4$ the number of free parameters in the model fitting, $a_{\rm 0}=$4.74\kms~mas$^{-1}$yr~kpc$^{-1}$ a conversion factor from a proper motion to a transverse velocity, and $D\equiv$2.0~kpc the distance to the maser source from the Sun as determined below. $\mu_{ix}$ and $\mu_{iy}$ are the observed proper motion components in the R.A. and decl. directions, respectively, $\sigma_{\mu_{ix}}$ and $\sigma_{\mu_{iy}}$ their uncertainties; $u_{iz}$ the observed LSR velocity, and $\sigma_{iz}$ its uncertainty. For simplicity we assume a radially expanding outflow. In this case, the modeled velocity vector, {\boldmath{$w_{i}$}} $(w_{ix},w_{iy},w_{iz})$, is given as 

\begin{equation}
\mbox{\boldmath $w_{i}$} =  
\mbox{\boldmath $V_{\rm 0}$}(V_{{\rm 0}x},V_{{\rm 0}y}, V_{{\rm 0}z})+
V_{\rm exp}(i)\frac{\mbox{\boldmath $r_{i}$}}{r_{i}}, \\
\label{eq:wi}
\end{equation}

\noindent 
where
\begin{eqnarray}
\mbox{\boldmath $r_{i}$} & = & \mbox{\boldmath $x_{i}$}(x_{i}, y_{i}, z_{i})
-\mbox{\boldmath $x_{\rm 0}$}(\Delta X_{\rm 0}, \Delta Y_{\rm 0}), \\
\label{eq:ri}
z_{i} & = & \frac{(u_{iz}-V_{{\rm 0}z})(r^{2}_{ix}+r^{2}_{iy})}
{(u_{ix}-V_{{\rm 0}x})r_{ix}+(u_{iy}-V_{{\rm 0}y})r_{iy}}, \\
\label{eq:zi}
u_{ix} & = & \mu_{ix}a_{\rm 0}D,\;\; u_{iy}=\mu_{iy}a_{\rm 0}D.
\label{eq:uix-uiy}
\end{eqnarray}

Table \ref{tab:kinematic-model} gives the parameters derived through fitting. A magenta ellipse in Figure \ref{fig:H2O}(a)--(c) indicates the estimated location of the originating point of the outflow; its size indicates the uncertainty of the location. The location appears biased to the eastern maser feature cluster due to the biased maser distribution and the complicated maser motions of the eastern cluster. The radial expansion velocities of the individual features are calculated after the model fitting using the following equation (Equation 7 of \citealt{ima11})

\begin{equation}
V_{\rm exp}(i)=\frac{(u_{ix}-V_{{\rm 0}x})r_{ix}+(u_{iy}-V_{{\rm 0}y})r_{iy}
+(u_{iz}-V_{{\rm 0}z})r_{iz}}{r_{i}}.
\label{eq:vexp}
\end{equation}

\begin{figure}
\epsscale{1.0}
\plotone{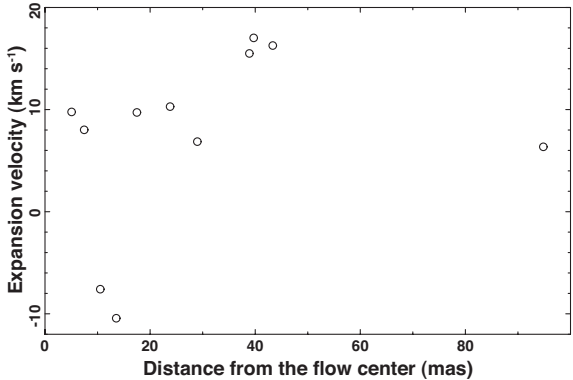}
\caption{Distribution of the expansion velocities of the individual \h2o maser features in the central region of \iras1846, which are derived from the fitting to a radially expanding flow model.\label{fig:spherical-flow}}
\end{figure}

\noindent
Figure \ref{fig:spherical-flow} shows the distribution of the expansion velocities. The data points seem to concentrate around an expansion velocity of $\approx$9\kms\ near the outflow origin and $\sim$17\kms\ at the outer region. It is difficult, however, to recognize such outward acceleration from the limited number of data points. These values are typical of those seen in other AGB CSEs.  

One can estimate the distance to I18460 using the {\it statistical parallax method}. This method assumes a uniform and random velocity field of gas clumps, therefore a velocity variance on the plane of the sky is expected to be equal to that in the line of sight. The equality of transverse and line-of-sight (or radial) velocity variances is also applicable to a spherically expanding model, such as the case of I18460. From the variances of proper motions and radial velocities of maser features, $\sigma^{2}_{\mu}$ and $\sigma^{2}_{v}$, respectively, the source distance and its uncertainty, $D$ and $\sigma_D$ respectively, are estimated using the following equations \citep{sch81}: 

\begin{equation}
D=\sigma_{v}/\sigma_{\mu},\;\;\; 
\frac{\sigma_{D}}{D}\sim
\left\{\frac{1}{2N_{m}}+\frac{1}{4N_{m}}\left[1-\left(\frac{\epsilon_{\rm obs}}{\sigma^{\rm obs}_{\mu}}
\right)^{2}\right]^{-1}\right\}^{1/2}. 
\label{eq:distance}
\end{equation}

\noindent 
We calculated a radial velocity variance, including a contribution from the expanding flow, $\sigma^{2}_{v_{z}}=$40.38~(km~s$^{-1}$)$^{2}$ from a mean square deviation from the systemic velocity. The obtained proper motion variance was corrected for its error, $\epsilon^{2}_{\rm obs}$,  by the following equation

\begin{equation}
\sigma^{2}_{\mu}=(\sigma^{\rm obs}_{\mu})^{2}-\epsilon^{2}_{\rm obs}.
\label{eq:correction}
\end{equation} 

\noindent
We obtained the corrected variances of proper motions, $(\sigma_{\mu_{x}},\sigma_{\mu_{y}}) =(0.61, 0.63)$ in units of mas~yr$^{-1}$. Their errors, (0.38, 0.39) (mas~yr$^{-1}$), were derived from the uncertainties of the measured proper motions. These values give statistical parallax distance values: 2.2$\pm$0.6~kpc (between $\sigma_{\mu_{x}}$ and $\sigma_{v_{z}}$) and 2.1$\pm$0.6~kpc (between $\sigma_{\mu_{y}}$ and $\sigma_{v_{z}}$), yielding a weighted mean value of $D=2.1\pm 0.6$~kpc. The justification of the derived distance is discussed in Section \ref{sec:distance}. 

\subsection{The Stellar Secular Motion}
\label{sec:secular-motion}

We obtained five {\it absolute} proper motions of \h2o\ maser features observed in 2008--2009. Because of intrinsic fading of the \h2o\ masers, we detect the central component with a smaller number of features than that in 2006--2007. The northeast--southwest alignment and the proper motions of the \h2o maser features found in 2008--2009 were similar to those found in 2006--2007 (Figure \ref{fig:H2O}(c) ). Therefore, we can assume that the masers from the different epochs are likely associated with a common origin. Two out of the five maser features were detected in all three epochs in 2008--2009. We fitted their absolute motions with respect to the position-reference source J1851 to the modeled motions, each of which was assumed to be composed of a constant velocity motion and an annual parallactic motion ($N_{\rm p}=5$). 

Table \ref{tab:secular-motion} shows the results of the maser motion fitting. We obtained annual parallax values of 0.45--0.58~mas, corresponding to a range of the distance to I18460 of  1.7--2.2~kpc. However, the time baseline was too short (only 5 months) and the degree of freedom in the model fitting was too low ($2N_{\rm e}-N_{\rm p}=1$ where $N_{\rm e}=3$ is the number of observing epochs) for reliable analysis. Therefore, the derived parallax was used just to extract the non-negligible parallax modulations and to obtain more precise linear proper motions required for further analysis. Thus we derived the absolute proper motion of the position-reference maser feature (\iras1846:I2013-{\it 19}) of  $(\mu_{\rm X}, \mu_{\rm Y})=(-3.73\pm1.17, -6.61\pm0.41)$  mas in the celestial coordinates. 

\begin{figure}
\epsscale{1.0}
\plotone{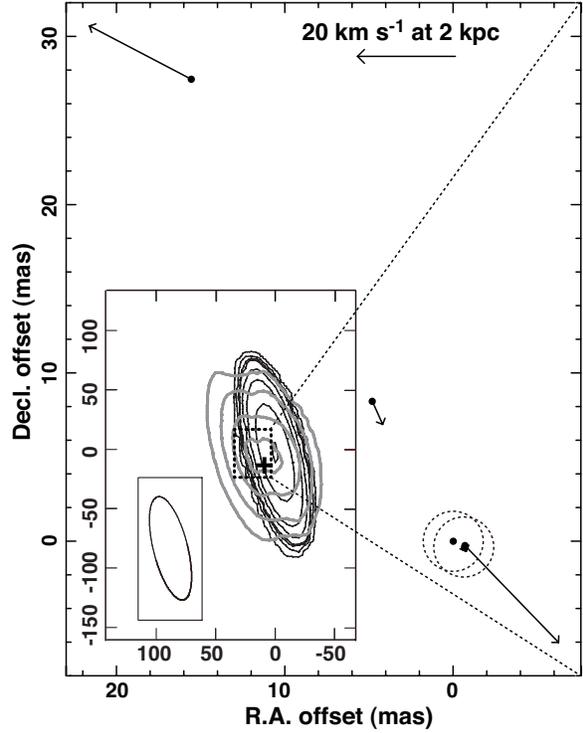}
\caption{\h2o and  OH masers observed in 2008 September--2009 February and in 2007 June, respectively. Here the self-calibration based map of the OH masers is displayed. The relative positions and proper motions of \h2o\ masers (filled circles and arrows, respectively) were measured with respect to the position-reference maser feature \iras1846:I2013-{\it 19}. The location of the two features \iras1846:I2013-{\it 18} and {\it 19}, whose absolute motions were measured, are indicated by the centers of the dotted circles. The left-bottom sub-panel shows the velocity-integrated brightness distribution of the  OH masers in the blue-shifted (black contours, \vlsr$=$111.0--115.6\kms) and the red-shifted (gray contours, \vlsr$=$138.6--140.1\kms) components, respectively. The contour levels are 0.18, 0.72, 1.44, 1.98, 5.76, 11.52, 23.04, and 39.60 Jy~beam$^{-1}$km~s$^{-1}$. The coordinate offsets are given with respect to the \vlsr$=$111.7\kms\ component of the OH masers. The ellipse at the left-bottom corner of this sub-panel shows the synthesized beam pattern. The plus symbol indicates the location of the \h2o maser feature \iras1846:I2013-{\it 19} in this coordinate system.
\label{fig:astrometry}}
\end{figure}

Figure \ref{fig:astrometry} shows the relative proper motions of the \h2o\ maser features found in 2008--2009 with respect to the position-reference feature. They exhibit expanding flow motions, similar to the masers shown in Figures \ref{fig:H2O}(b)--(c) . This is clear even if we eliminate the motion of the feature, \iras1846:I2013-{\it 17}, which was detected at only two epochs and whose measured large relative proper motion was doubtful. Using the proper motion data of the features \iras1846:I2013- {\it 16}, {\it 18}, {\it 19}, and {\it 20}, we estimated the stellar motion to be $(\Delta_{\mu X}^{\ast}, \Delta_{\mu Y}^{\ast})\approx (0.49, 0.14)$(mas~yr$^{-1}$) with respect to the position-reference feature. This corresponds to an expanding velocity of the CSE of $\approx$5\kms. The uncertainty of the relative stellar motion may be comparable to or smaller than this value. We added, therefore, this stellar motion to both the absolute proper motion of the reference feature and its error to obtain a stellar secular motion, $(\mu_{\rm X}, \mu_{\rm Y})=(-3.24\pm1.27, -6.47\pm0.43)$  mas. The absolute coordinates of the reference feature were determined to be 
$\alpha_{\rm J2000}=$18$^{\rm h}$48$^{\rm m}$43$^{\rm s}$\hspace{-2pt}.02792$\pm$0$^{\rm s}$\hspace{-2pt}.00006, 
$\delta_{\rm J2000}=$~$-$01$^{\circ}$48$^{\prime}$30\farcs 46141$\pm$0\farcs 00019 
in 2008 September. 

\subsection{The Circumstellar Envelope Traced by OH Masers}
\label{sec:OH}

Figure \ref{fig:astrometry} shows the velocity-integrated brightness contour map of the two velocity components of the OH masers obtained from the self-calibration-based image cube. The origin of the contour map is set to the position of the reference OH maser spot at \vlsr$=$111.7\kms, whose absolute coordinates were determined on the phase-referencing based image cube to be
$\alpha_{\rm J2000}=$18$^{\rm h}$48$^{\rm m}$43$^{\rm s}$\hspace{-2pt}.02701$\pm$0$^{\rm s}$\hspace{-2pt}.00007, 
$\delta_{\rm J2000}=$~$-$01$^{\circ}$48$^{\prime}$30\farcs 4526$\pm$0\farcs 0026. Interestingly, the position-reference \h2o\ feature and OH spot were located within 10~mas (see Section \ref{sec:secular-motion}), corresponding to a linear scale of $\approx$20~AU, even taking into account the maser positions at the different epochs and the maser motions in the period of these epochs. The OH masers and the central \h2o\ ones have a common LSR velocity range (110--140\kms) and indicate similar expansion velocities (10--20\kms, Section \ref{sec:center}). It is strongly suggested that these masers share a common origin. 

\begin{figure}
\epsscale{1.0}
\plotone{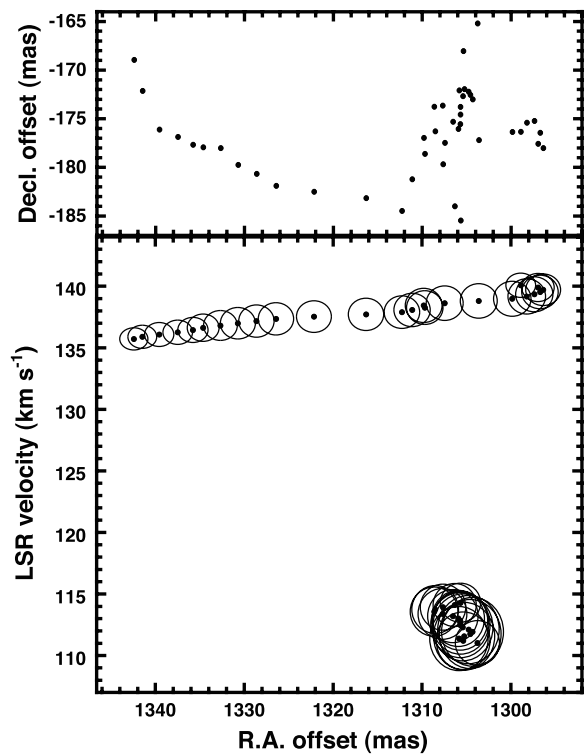}
\caption{Angular distribution and position--velocity diagrams of the OH masers \iras1846, which were detected on the phase-referencing-based image cube, denoted by filled circles. The size of the circle around the maser in the position--velocity diagram is proportional to logarithmic scale of its intensity. \label{fig:OH}}
\end{figure}

Figure \ref{fig:OH} shows the distribution of the OH masers in more detail. Each of the maser spots was resolved out on baselines to the Hartebeesthoek station: the spot size was estimated to be $\gtrsim$25~mas. The blue-shifted and red-shifted clusters of OH maser spots were split into the east--west direction by a separation of $\approx$50~mas, comparable to the extent of the central component of \h2o\ masers. The origin of a systematic LSR velocity gradient visible in only the red-shifted component is unknown and it should be investigated whether this velocity gradient is always visible. But we note that the directions of the elongation of the \h2o maser distribution (Section \ref{sec:center}), the cluster splitting, and the velocity gradient are perpendicular to the axis of the \h2o\ maser jet (Section \ref{sec:jet}). Furthermore, the expansion speeds and the angular distribution sizes are equal for the central component of \h2o masers and the OH masers. These properties are similar to those of \h2o\ and OH masers in W~43A \citep{ima02a, ima07a, ima08}, where the existence of \h2o\ masers that are not associated with the collimated jet was first claimed. However, the kinematics of the W43A \h2o\ masers looks different from that of the OH maser although the kinematic information was quite limited in that source. The central component of \h2o\ masers in I18460 is the first example whose spatiokinematical structure was determined through the construction of the radially expanding flow model as described in Section \ref{sec:center} and compared with the kinematics of the OH masers.  

\section{DISCUSSION}
\label{sec:discussion}

\subsection{Distance to \iras1846}
\label{sec:distance}

The results of this paper favor a much shorter distance to I18460 ($D\sim$2.0~kpc) than the derived kinematic distance ($D_{\rm kin}\sim $6.9~kpc). The latter value is obtained when one uses only the radial velocity and assumes a circular orbit of I18460 in the Milky Way. This suggests that this source is located close to a tangential point in the Milky Way. Using both the radial velocity and the proper motion (Section  \ref{sec:secular-motion}), $D_{\rm kin}\sim$6.0 or 7.8~kpc is obtained. However, unless the maser source is of a young generation, the assumption of a circular orbit is unlikely (see also Section \ref{sec:secular-motions2}). 

As described in Section \ref{sec:center}, the statistical parallax distance to I18460 is estimated as $D_{\rm stat}=2.1\pm 0.6$~kpc. As shown in Figure \ref{fig:H2O}, the distribution of the central components of \h2o\ masers is elongated in the east--west direction. In addition, as shown in Figure \ref{fig:spherical-flow}, the expanding flow appears to be accelerated outward. These results may affect this estimate of distance; the variance of proper motions may have a bias in the east--west direction. In practice, as already mentioned, the outward acceleration is negligible and the calculated results support that such a variance bias is also negligible. On the other hand, if these \h2o\ masers are associated with a face-on equatorial flow, then a transverse velocity variance larger than that in the radial direction is expected,  causing a possible bias to a shorter distance. This possibility is also unlikely when taking into account the similarity of the kinematics of the \h2o\ masers to that of the 1612~MHz OH masers, which suggest spherical expansion. Thus, even the possible bias in $D_{\rm stat}$ cannot explain the much shorter distance to I18460. The annual parallax unambiguously rules out a large $D_{\rm kin}$, although the uncertainty of the parallax is very large. 

\subsection{The Collimated Jet in \iras1846}
\label{sec:jet2}

The driving mechanism of the WF jet is still in debate, but a comparison of the jet parameters with other parameters of the WF sources will shed light on the mechanism. For those WFs with an estimated three-dimensional jet velocity, however, we find so far no clear correlation among the three-dimensional velocity, dynamical age, and degree of collimation of the jet and the stellar luminosity (roughly evaluated with the {\it IRAS} flux at the 25~$\mu$m band multiplied by the square of the source distance). 

An offset of the center velocity of a collimated jet from the systemic velocity of a CSE and an angular offset of the originating point of a collimated jet from that of a CSE are observable in the maser spatiokinematics. The existence of these offsets implies the existence of a binary system in which the jet and the CSE originate from different host stars. In the case of I18460, as shown in Figures \ref{fig:H2O} and \ref{fig:astrometry} (see Sections \ref{sec:jet}, \ref{sec:center}, and \ref{sec:OH}), an upper limit to the angular offset mentioned above was given to be $\lesssim$20~mas ($\lesssim$40~AU at 2~kpc). The center velocity offset was derived to be 10--40\kms\ from the center velocity of the jet \h2o\ masers and the systemic velocity derived from the OH maser spectrum (see Section \ref{sec:jet}). The derived velocity offset is roughly consistent with those in IRAS~16342$-$3814 and IRAS~18286$-$0959, in which the center velocities of \h2o\ and OH masers differ by up to 30\kms (\citealt{cla09, yun11, ima13a}, hereafter Paper I). However, uncertainties of these center velocities of the jet masers are large. Unless the point symmetry of the \h2o\ maser spectra is well examined, the large uncertainty of the center velocity of the jet cannot be reduced, so it prevents us from any realistic constraint on the orbital parameters of a supposed binary system or being convinced of the existence of the binary system. 

Here, we consider the possibility of identification of a binary system, with one component driving the WF jet while the other hosts the CSE. The latter may be an O-rich intermediate-mass AGB or post-AGB star \citep{ima12a}, {\bf so here we assume a current stellar mass of $\sim$1~$M_{\sun}$}. In the case of W~43A, in which the center velocities of the \h2o, OH, and SiO masers are estimated to be consistent with one another within a few \kms, it is possible to consider a separation of 100~AU between two stars, which is comparable to the upper limit to the separation estimated from the maser observations (\citealt{ima02a, ima05}, see also the case of OH~12.8$-$0.9, \citealt{bob07}). If a companion orbits the supposed star with a velocity of 10--40\kms\ as estimated in I18460, the separation of these stars may be in a range of 0.6--9~AU, which cannot be spatially resolved with the current analysis technique relying on the maser data. On the other hand, the case of the highest velocity offset corresponds to a velocity fluctuation of the host star of $\gtrsim$0.3\kms\ at 9~AU and an orbital period of $\lesssim$30~yr. These derived parameters will be examined in future infrared and sub-millimeter observations which can resolve such a scale and which will enable us to conduct high-resolution spectroscopy of molecular lines associated with the central stars and the accompanying jets and CSEs. In any case, finding point a symmetric distribution of the jet masers and a reasonable jet model may be necessary to more precisely estimate the center velocity of the jet, which is essential to identify a binary system.

It is noteworthy that the \h2o\ masers in I18460 had faded between the 2006--2007 and 2008--2009 seasons. It should be taken into account that stellar \h2o\ masers are highly variable in both their angular and velocity distribution and some of the variation will be correlated with stellar pulsation (e.g., \citealt{bow94}). In the case of the WF jet masers, however, it is  possible that we are observing {\it the evolution} of the jet on a timescale comparable to the estimated dynamical time scale (see Section \ref{sec:jet}). We have to confirm, using future long-term observations, whether we are observing the true ``death" of a WF or recurrent jet ejections; this should reveal the role of the WF jet in shaping the future planetary nebula. 

\subsection{The Relic AGB CSE in \iras1846}
\label{sec:equatorial-flow}

The AGB CSE/equatorial flow as well as the collimated fast jet is important components to probe the evolutionary status of the WF source. As mentioned in Section \ref{sec:OH}, the OH masers and the central component of \h2o\ masers in I18460 resemble those of AGB CSEs. Taking into account that these masers do not have a high expansion velocity ($V_{\rm exp}\approx$8--15\kms) and they trace only a limited volume of the CSE or equatorial flow, these masers likely have close association with each other as shown in some OH/IR stars. 

It is noteworthy that the equatorial elongation in the \h2o/OH CSE is still marginal. It is suggested that the 1612~MHz OH masers are radially amplified and it is possible that their area projected on the plane of sky is sometimes smaller than that of \h2o\ masers. Even taking into account these factors, the noted properties of the OH and \h2o\ masers in I18460 imply that this CSE may be a relic AGB CSE rather than a product of recent rapid growth of an equatorial torus or flow as suggested by \citet{hug07}. Because the collimated jet of I18460 still looks extremely young (it has a dynamical age of only $\sim$6~yr, see Section \ref{sec:jet}) it is too early to observe the intrinsic evolution of the equatorial flow even if it is recently launched. \citet{hug07} suggested a time lag of a few hundred years between the rapid growth of the equatorial torus and the ignition of the collimated jet. In practice, the dynamical age of the I18460 CSE is only $\approx$140~yr. Any dynamical age derived from the data of \h2o\ and OH masers may be much shorter than the true age of the dynamical component (i.e., jet or CSE). Nevertheless we believe that the former time scale is meaningful when comparing the dynamical time scales among the maser sources when discussing the evolution of the maser sources. In fact, \citet{ima05} and \citet{bob05} have found the temporal evolution of the \h2o\ maser jets in W~43A and OH~12.8$-$0.9 on the decade scale, respectively. In both cases, the expansion rate of the \h2o\ maser region in the jet is roughly consistent with that expected from the dynamical ages.\footnote
{Here we note a recent unexpected result for IRAS~16342$-$3814 \citep{cla12}, in which relative proper motions of the \h2o\ observed in 2008--2009 were completely different from those in 2002 \citep{cla09}. The authors pointed out two possible causes of the result. We also consider their second possibility as an extreme case, in which \h2o\ masers reached the edge of the CSE and maser excitation does not occur at a distance larger than that to this point from the star.} 

It is unclear whether the existence of the relic AGB CSE indicates that its hosting source is in the early phase of the WF. As already mentioned, the dynamical ages of the CSEs are much longer than those of the jets. The jet will affect only the inner part of the CSE in the WF phase, and the volume of the interaction with the highly collimated jet will be still quite limited. Therefore, it is difficult to find any correlation between the kinematic parameters of the jet and the CSE with the overall properties of the CSE as traced by its infrared colors. However, if we suppose a beaming effect of \h2o\ and OH masers, the small interaction with the jet may enhance the difference in preferred locations of these masers in the CSE. Similar to 1612~MHz OH masers, \h2o\ masers in a CSE prefer radial beaming, in which the maser emission is amplified radially with respect to the central star (e.g., \citealt{tak94}). In this case, one can see \h2o\ masers associated with the CSE if the collimated jet breaks up the CSE only along the plane of the sky while THE CSE is still maintained in the line of sight. After checking the previous VLBI observations of WFs \citep{ima02a,bob07,cla09,day10,yun11}, we speculate that the central \h2o\ masers of the WFs, associated with relic AGB CSEs, may be visible in the case in which the jet inclination angle with respect to the line of sight is sufficiently large ($i>$30\arcdeg). This speculation should be further examined in future VLBI observations of other WFs.  

\subsection{The Secular Motion of \iras1846}
\label{sec:secular-motions2}

As shown in Sections \ref{sec:center} and \ref{sec:secular-motion}, we have derived the distance and the secular motion of I18460 to be $D=2.1\pm0.6$~kpc and  $(\mu_{\rm X}, \mu_{\rm Y})=(-3.24\pm1.27, -6.47\pm0.43)$  mas, respectively. Based on these values, we further estimate the location and the three-dimensional motion of I18460 in the Milky Way. Table \ref{tab:MW-motion} gives the list of relevant parameters. We note common properties of the WFs whose kinematics in the Milky Way have been estimated (IRAS~19134$+$2131, \citealt{ima07b}; IRAS~18286$-$0959, Paper I). Although some of the WFs are located several hundred parsecs away from the Galactic mid-plane (IRAS~163424$-$3814, IRAS~19134$+$2131), a large fraction of the WFs is located in the Galactic thin disk as seen for massive-star forming regions. Nevertheless, their three-dimensional motions have large deviations by $\gtrsim$50\kms\ from those expected from the circular Galactic rotation.\footnote
{We are convinced that our program correctly calculates three-dimensional vectors of maser source motions on the Galactic plane, whose results have been checked by comparing with those in previous publications of maser source astrometry by other authors (e.g., \citealt{hon13}).}  It is expected that the motions of the WF sources have such large deviation as a result of dynamical relaxation during their relatively long lifetimes ($>10^{7}$~yr).  However, this is inconsistent with the observation that some of the WFs, including I18460, still persist in the Galactic mid-plane. Paper I point out the possibility of a binary motion for such a peculiar motion.

\section{CONCLUSIONS}

The present VLBA and EVN observations have revealed the spatiokinematical structures of both of the collimated fast jet and the slowly expanding equatorial flow or CSE associated with I18460. The dynamical ages of these kinematical components are extremely short ($\sim$6 and $\sim$140~yr, respectively) although these values may not be equal to their true ages. The suggestion that I18460 should be in a very young stage of the WF evolution is supported by its association with the CSE as seen in OH/IR stars. The driving sources of the different components are located within $\sim$10~mas ($\sim$20~AU), a situation very similar to the case of the \h2o\ and SiO masers in W~43A \citep{ima05}. By adding the present results, the common properties of the three-dimensional velocity vectors of the WFs in the Milky Way have been elucidated, which implies that the WFs harbor binary systems. Future observations that enable us to resolve such binary systems ($\lesssim$20~AU at a few kilo parsecs), as those conducted with new facilities such as the Atacama Large Millimeter-submillimeter Array (ALMA), will directly shed light on the evolution of AGB/post-AGB stars generating the WFs. 

\acknowledgments
The VLBA/National Radio Astronomy Observatory is a facility of the National Science Foundation, operated under a cooperative agreement by Associated Universities, Inc. The European VLBI Network (EVN) is a joint facility of European, Chinese, South African and other radio astronomy institutes funded by their national research councils. HI has been supported financially by Grant-in-Aid for Young Scientists (B) from the Ministry of Education, Culture, Sports, Science, and Technology (18740109) and for stay at ICRAR by the Strategic Young Researcher Overseas Visits Program for Accelerating Brain Circulation funded by the Japan Society for the Promotion of Science (JSPS). HI and SD have been financially supported by Grant-in-Aid for Scientific Research from JSPS (20540234). JN has been supported by the Research Grants Council of Hong Kong (project code: HKU 703308P, HKU 704209P, HKU 704710P, and HKU 704411P), and the Small Project Funding of The University of Hong Kong (201007176004). 

%%%%% References

\clearpage
%%%%% Table 1 %%%%%%%
\begin{table}[h]
\caption{Parameters of the VLBA Observations and Data Reduction for Individual Epochs}\label{tab:status}

\scriptsize
\begin{tabular}{l@{ }c@{ }ccc@{ }l@{ }r} \hline \hline
Observation & Epoch & $V_{\mbox{\scriptsize ref}}$ \tablenotemark{1} & Noise\tablenotemark{2} & Dynamic 
& \multicolumn{1}{c}{Beam\tablenotemark{4}}
& \\
code & (yy/mm/dd) & (km~s$^{-1}$)
& (mJy beam$^{-1}$) & Range\tablenotemark{3} 
&  \multicolumn{1}{c}{(mas)} & $N_{\mbox{\scriptsize f}}$\tablenotemark{5} 
\\ \hline 
BI033A \dotfill & 06/09/15 & 117.7 & 2.6 & $\sim$90
& 1.31$\times$0.88, 2$^{\circ}$\hspace{-2pt}.7 & 34 \\
BI033B \dotfill & 06/11/09 & 117.6 & 3.1 &$\sim$220 
& 1.37$\times$0.96, 17$^{\circ}$\hspace{-2pt}.4 & 27 \\
BI033C \dotfill & 07/02/03 & 118.3 & 1.9 &$\sim$110  
& 1.50$\times$0.96, $-$12$^{\circ}$\hspace{-2pt}.7 & 16 \\ \hline
BI036A \dotfill & 08/09/05 & ---\tablenotemark{6} & 12  &$\sim$30
& 1.77$\times$1.18, 11$^{\circ}$\hspace{-2pt}.1 & 11 \\
BI036B \dotfill & 08/11/18 & ---\tablenotemark{6} & 11 &$\sim$50 
& 1.65$\times$0.92, $-$2$^{\circ}$\hspace{-2pt}.1 & 10 \\ 
BI036C \dotfill & 09/02/06 &---\tablenotemark{6} &  9.7 &$\sim$40
& 1.58$\times$0.95, $-$3$^{\circ}$\hspace{-2pt}.5 & 3 \\ \hline
\end{tabular}

\tablenotetext{1}{ Local-standard-of-rest (LSR) velocity at the phase-reference spectral channel.} 
\tablenotetext{2}{ Root-mean-squares noise in the emission-free spectral channel image.} 
\tablenotetext{3}{ Achievable best dynamic range on the spectral channel image.} 
\tablenotetext{4}{ Synthesized beam size resulting from natural weighted visibilities; major and minor full-widths at half of maximum and position angle.}
\tablenotetext{5}{ Number of the detected maser features.}
\tablenotetext{6}{ Position- and phase-reference was J185146.7$+$003532.}  
\end{table}

%%%%% Table 2 %%%%%
\begin{deluxetable}{r@{ }r@{ $\pm$}r@{  }r@{ $\pm$}rc@{ }c@{ }}
%\rotate
\tablecaption
{Parameters of the Detected \h2o\ Maser Features Found in 2006--2007 \label{tab:features_h2o_a}}
\tabletypesize{\scriptsize}
\tablecolumns{8}
\tablehead{$V_{\mbox{\tiny LSR}}$\tablenotemark{1} 
& \multicolumn{2}{c}{R.A. Offset\tablenotemark{2}} & 
\multicolumn{2}{c}{Decl. Offset\tablenotemark{2}} & 
$I$\tablenotemark{3}   & \multicolumn{1}{c}{$\Delta V$\tablenotemark{4}} \\
(km~s$^{-1}$) & \multicolumn{2}{c}{(mas)} &  \multicolumn{2}{c}{(mas)} 
& (Jy~beam$^{-1}$)& (km~s$^{-1}$)} 
\startdata
\multicolumn{7}{c}{2006 September 15} \\ \hline
$  241.79$&$   -31.66$&$0.02$&$   -52.66$&$0.06$&$    0.03$&$ 0.84$\\
$  132.14$&$   -31.60$&$0.02$&$    15.24$&$0.04$&$    0.07$&$ 0.84$\\
$  130.56$&$   -30.76$&$0.02$&$    16.96$&$0.02$&$    0.09$&$ 0.84$\\
$  123.43$&$   -21.86$&$0.02$&$   -16.52$&$0.02$&$    0.15$&$ 1.05$\\
$  121.26$&$   -12.40$&$0.01$&$   -16.56$&$0.02$&$    0.36$&$ 0.63$\\
$  121.26$&$   -12.59$&$0.02$&$   -15.66$&$0.01$&$    0.06$&$ 0.21$\\
$  121.26$&$   -12.15$&$0.01$&$   -17.44$&$0.01$&$    0.09$&$ 0.21$\\
$  118.95$&$    -3.94$&$0.03$&$    -6.99$&$0.05$&$    0.31$&$ 0.63$\\
$  118.87$&$    -5.68$&$0.02$&$     6.25$&$0.06$&$    0.43$&$ 0.84$\\
$  118.37$&$    -0.26$&$0.02$&$     3.57$&$0.02$&$    2.39$&$ 1.89$\\
$  119.16$&$    -4.34$&$0.11$&$    13.37$&$0.08$&$    0.19$&$ 0.84$\\
$  118.74$&$    -4.67$&$0.02$&$    -3.71$&$0.04$&$    0.35$&$ 0.21$\\
$  118.10$&$    -0.11$&$0.01$&$    -0.85$&$0.05$&$    1.54$&$ 0.63$\\
$  117.85$&$    -0.01$&$0.02$&$    -0.01$&$0.05$&$    2.66$&$ 1.90$\\
$  117.89$&$    -0.01$&$0.05$&$    -1.08$&$0.05$&$    0.90$&$ 0.84$\\
$  116.94$&$    -0.61$&$0.01$&$    -2.70$&$0.04$&$    0.09$&$ 0.84$\\
$  113.54$&$   -61.82$&$0.03$&$    33.49$&$0.03$&$    0.07$&$ 1.47$\\
$  113.68$&$   -61.59$&$0.01$&$    32.56$&$0.01$&$    0.06$&$ 0.42$\\
$  113.47$&$   -62.01$&$0.01$&$    34.34$&$0.01$&$    0.12$&$ 0.21$\\
$  -17.11$&$     1.24$&$0.05$&$    60.68$&$0.09$&$    0.04$&$ 1.26$\\
$  -17.49$&$     1.22$&$0.11$&$    60.73$&$0.18$&$    0.13$&$ 0.84$\\
$  -19.52$&$     1.24$&$0.02$&$    60.74$&$0.06$&$    0.12$&$ 2.32$\\
$  -17.82$&$     1.79$&$0.05$&$    59.66$&$0.10$&$    0.04$&$ 1.47$\\
$  -19.71$&$     1.21$&$0.02$&$    60.80$&$0.03$&$    0.13$&$ 0.63$\\
$  -22.96$&$    -3.40$&$0.03$&$    65.86$&$0.06$&$    0.04$&$ 0.21$\\
$  -43.20$&$    -2.21$&$0.04$&$    60.06$&$0.11$&$    0.04$&$ 0.84$\\
$  -47.56$&$    -1.09$&$0.01$&$    60.41$&$0.02$&$    0.40$&$ 3.16$\\
$  -46.99$&$    -0.86$&$0.01$&$    59.46$&$0.01$&$    0.11$&$ 1.26$\\
$  -49.15$&$    -0.14$&$0.01$&$    60.82$&$0.01$&$    1.38$&$ 2.32$\\
$  -49.22$&$     0.09$&$0.01$&$    59.86$&$0.01$&$    0.27$&$ 0.84$\\
$  -49.31$&$    -0.30$&$0.02$&$    61.73$&$0.01$&$    0.17$&$ 0.21$\\
$  -50.65$&$    -1.32$&$0.01$&$    60.22$&$0.02$&$    0.24$&$ 1.68$\\
$  -54.21$&$     0.12$&$0.02$&$    60.74$&$0.06$&$    0.08$&$ 2.11$\\
$  -67.38$&$     1.64$&$0.01$&$    59.34$&$0.03$&$    0.18$&$ 1.05$\\
\hline
\multicolumn{7}{c}{2006 November 9} \\ \hline
$  133.06$&$   -30.57$&$0.02$&$    19.09$&$0.05$&$    0.03$&$ 0.42$\\
$  132.08$&$   -31.84$&$0.02$&$    15.37$&$0.02$&$    0.11$&$ 1.26$\\
$  130.52$&$   -30.99$&$0.01$&$    17.08$&$0.01$&$    0.15$&$ 1.05$\\
$  130.96$&$   -41.48$&$0.02$&$    -7.30$&$0.05$&$    0.03$&$ 0.11$\\
$  123.49$&$   -22.03$&$0.01$&$   -16.59$&$0.02$&$    0.13$&$ 0.84$\\
$  121.26$&$    16.93$&$0.02$&$     4.74$&$0.03$&$    0.08$&$ 0.63$\\
$  121.26$&$   -12.48$&$0.01$&$   -16.69$&$0.01$&$    0.19$&$ 0.42$\\
$  119.58$&$    -4.45$&$0.03$&$    -9.40$&$0.03$&$    0.18$&$ 0.11$\\
$  118.96$&$    -5.72$&$0.02$&$     6.12$&$0.05$&$    0.91$&$ 1.05$\\
$  118.34$&$    -0.23$&$0.01$&$     3.57$&$0.05$&$    3.14$&$ 1.69$\\
$  117.26$&$    -0.21$&$0.01$&$     3.23$&$0.04$&$    0.14$&$ 0.42$\\
$  118.26$&$    -0.09$&$0.01$&$    -0.71$&$0.06$&$    2.08$&$ 0.63$\\
$  118.02$&$    -0.07$&$0.05$&$    -1.03$&$0.10$&$    0.70$&$ 1.05$\\
$  118.32$&$    -0.02$&$0.01$&$     0.31$&$0.01$&$    0.69$&$ 0.11$\\
$  118.32$&$    -0.05$&$0.01$&$     2.66$&$0.01$&$    0.64$&$ 0.42$\\
$  118.10$&$     0.04$&$0.01$&$    -1.64$&$0.07$&$    0.68$&$ 0.42$\\
$  118.10$&$    -0.03$&$0.01$&$    -0.37$&$0.01$&$    2.07$&$ 0.11$\\
$  117.87$&$    -0.01$&$0.02$&$    -0.06$&$0.06$&$    2.40$&$ 0.63$\\
$  117.81$&$     0.12$&$0.03$&$    -0.84$&$0.01$&$    1.07$&$ 1.26$\\
$  117.89$&$    -0.12$&$0.01$&$     0.85$&$0.01$&$    1.18$&$ 0.11$\\
$  117.57$&$     0.01$&$0.01$&$     0.05$&$0.01$&$    1.22$&$ 1.26$\\
$  116.85$&$    -0.59$&$0.01$&$    -2.76$&$0.04$&$    0.15$&$ 1.05$\\
$  -17.74$&$     1.54$&$0.02$&$    62.05$&$0.03$&$    0.13$&$ 1.90$\\
$  -19.87$&$     1.44$&$0.05$&$    62.22$&$0.06$&$    0.15$&$ 2.32$\\
$  -21.07$&$     1.39$&$0.03$&$    62.23$&$0.05$&$    0.09$&$ 1.48$\\
$  -49.03$&$     0.07$&$0.01$&$    62.30$&$0.02$&$    0.87$&$ 2.53$\\
$  -48.89$&$     0.24$&$0.02$&$    61.36$&$0.04$&$    0.10$&$ 1.05$\\
\hline  
\multicolumn{7}{c}{2007 February 3} \\ \hline
$  133.27$&$   -30.64$&$0.02$&$    15.65$&$0.10$&$    0.02$&$ 0.63$\\
$  132.05$&$   -31.91$&$0.01$&$    12.00$&$0.02$&$    0.14$&$ 1.26$\\
$  130.55$&$   -31.08$&$0.02$&$    13.64$&$0.01$&$    0.10$&$ 1.05$\\
$  124.06$&$   -11.10$&$0.03$&$   -21.62$&$0.04$&$    0.03$&$ 0.63$\\
$  120.42$&$    10.11$&$0.04$&$    12.15$&$0.07$&$    0.03$&$ 0.11$\\
$  118.40$&$     0.04$&$0.01$&$    -0.02$&$0.08$&$    2.75$&$ 2.11$\\
$  118.29$&$     0.19$&$0.02$&$    -4.20$&$0.09$&$    1.25$&$ 0.63$\\
$  117.89$&$    -2.40$&$0.04$&$   -10.55$&$0.07$&$    0.07$&$ 0.11$\\
$  117.89$&$     0.26$&$0.01$&$    -3.80$&$0.16$&$    0.77$&$ 1.05$\\
$  117.68$&$    -3.36$&$0.04$&$   -12.56$&$0.08$&$    0.04$&$ 0.11$\\
$  116.65$&$    -0.31$&$0.02$&$    -6.31$&$0.04$&$    0.19$&$ 1.26$\\
$  -15.81$&$     2.66$&$0.11$&$    60.17$&$0.07$&$    0.63$&$ 6.32$\\
$  -45.36$&$     0.52$&$0.03$&$    61.29$&$0.07$&$    0.04$&$ 1.26$\\
$  -47.01$&$     0.50$&$0.02$&$    61.29$&$0.04$&$    0.06$&$ 1.48$\\
$  -48.50$&$     0.52$&$0.01$&$    61.26$&$0.04$&$    0.10$&$ 1.68$\\
$  -46.57$&$    -1.12$&$0.08$&$    60.76$&$0.08$&$    0.02$&$ 0.42$\\
\enddata

\tablenotetext{1}{ LSR velocity at the intensity peak.}
\tablenotetext{2}{ Position offset with respect to the phase-reference maser spot in the maser feature 
: \iras1846:I2013-{\it 6}.}
\tablenotetext{3}{ Peak intensity of the feature.}
\tablenotetext{4}{ Full velocity width of maser emission. 
The minimum is equal to the velocity spacing of a spectral channel (0.21\kms).}
\end{deluxetable}

\clearpage
%%%%% Table 3 %%%%%
%\rotate

\begin{deluxetable}{r@{ }r@{ $\pm$}r@{  }r@{ $\pm$}rc@{ }c@{ }}
\tablecaption{Same as Table \ref{tab:features_h2o_a} But in 2008--2009 \label{tab:features_h2o_b}}
\tabletypesize{\scriptsize}
\tablecolumns{7}
\tablehead{$V_{\mbox{\tiny LSR}}$\tablenotemark{1} & \multicolumn{2}{c}{R.A. offset\tablenotemark{2}} & 
\multicolumn{2}{c}{Decl. offset\tablenotemark{2}} & $I$\tablenotemark{3}  
& \multicolumn{1}{c}{$\Delta V$\tablenotemark{4}} \\
(km~s$^{-1}$) & \multicolumn{2}{c}{(mas)} &  \multicolumn{2}{c}{(mas)} & (Jy~beam$^{-1}$)& (km~s$^{-1}$) \\ \hline
\multicolumn{7}{c}{2008 September 5}}
\startdata
$   121.47$&$    39.52$& 0.06 &$    20.92$& 0.15 &        0.40 &   0.63 \\
$   121.47$&$    36.48$& 0.07 &$     0.72$& 0.11 &        0.44 &   0.63 \\
$   121.47$&$    36.86$& 0.07 &$    24.97$& 0.13 &        0.40 &   0.42 \\
$   121.04$&$    29.36$& 0.07 &$    13.39$& 0.12 &        0.72 &   1.05 \\
$   118.96$&$    13.82$& 0.14 &$   -14.06$& 0.19 &        1.02 &   1.05 \\
$   118.10$&$    13.13$& 0.10 &$   -14.46$& 0.30 &        2.75 &   0.84 \\
$   118.74$&$    13.63$& 0.10 &$   -15.26$& 0.07 &        0.66 &   0.84 \\
$   118.99$&$    17.03$& 0.09 &$     6.57$& 0.14 &        0.52 &   0.63 \\
$   117.89$&$    13.11$& 0.13 &$   -14.33$& 0.11 &        2.29 &   0.63 \\
$   117.26$&$    21.62$& 0.08 &$    14.57$& 0.27 &        0.52 &   0.63 \\
$   117.05$&$    18.62$& 0.06 &$    -5.74$& 0.11 &        0.54 &   0.42 \\
\hline \multicolumn{7}{c}{2008 November 18} \\ \hline
$   122.53$&$    -6.16$& 0.06 &$   -23.63$& 0.11 &        0.25 &   0.42 \\
$   121.12$&$    29.37$& 0.06 &$    11.56$& 0.05 &        0.62 &   1.47 \\
$   119.13$&$    13.38$& 0.12 &$   -16.12$& 0.09 &        3.30 &   1.69 \\
$   118.08$&$    12.68$& 0.08 &$   -16.41$& 0.13 &        2.33 &   0.63 \\
$   119.16$&$    13.41$& 0.10 &$     6.95$& 0.13 &        0.74 &   0.63 \\
$   119.37$&$    11.62$& 0.06 &$   -29.24$& 0.08 &        0.47 &   0.42 \\
$   119.11$&$    13.67$& 0.11 &$     7.26$& 0.16 &        0.74 &   0.84 \\
$   117.68$&$    12.26$& 0.16 &$   -16.82$& 0.12 &        1.18 &   0.42 \\
$   117.00$&$    18.14$& 0.04 &$    -7.91$& 0.13 &        0.81 &   1.05 \\
$   117.05$&$    18.48$& 0.06 &$    16.06$& 0.11 &        0.62 &   0.84 \\
\hline \multicolumn{7}{c}{2009 February 6} \\ \hline
$   119.26$&$    13.15$& 0.11 &$   -17.35$& 0.05 &        4.78 &   1.69 \\
$   118.27$&$    12.47$& 0.41 &$   -17.76$& 0.27 &        1.50 &   1.68 \\
$   118.74$&$    13.89$& 0.06 &$   -16.81$& 0.04 &        1.06 &   0.42 \\
\enddata

\tablenotetext{1}{ LSR velocity at the intensity peak.}
\tablenotetext{2}{ Position offset with respect to the phase-tracking center of \iras1846.}
\tablenotetext{3}{ Peak intensity of the feature.}
\tablenotetext{4}{ Full velocity width of maser emission. 
The minimum is equal to the velocity spacing of a spectral channel (0.21\kms).}
\end{deluxetable}

%%%%%%%%%%% Table 4 %%%%%%%%%%%%%%%%%%%
\begin{table*}[h]
\caption{Parameters of \h2o\ Maser Features Measured Their Relative Proper Motions in 2006--2007 and 2008--2009} 
\label{tab:pmotions}

\scriptsize
\begin{tabular}{lr@{ }rr@{ }rr@{ }rr@{ }rrrr} 
\hline \hline 
Maser Feature No.\tablenotemark{1} & \multicolumn{2}{c}{Position Offset\tablenotemark{2}}                         
 & \multicolumn{4}{c}{Proper Motion\tablenotemark{3}}
& \multicolumn{2}{c}{Radial Motion\tablenotemark{4}} 
& \multicolumn{3}{c}{Peak Intensity} \\    
 & \multicolumn{2}{c}{(mas)}
& \multicolumn{4}{c}{(\masyr)}
 & \multicolumn{2}{c}{(km s$^{-1}$)} & \multicolumn{3}{c}{(Jy~beam$^{-1}$)} \\   
(\iras1846:                 
 & \multicolumn{2}{c}{\ \hrulefill \ } 
 & \multicolumn{4}{c}{\ \hrulefill \ } 
& \multicolumn{2}{c}{\ \hrulefill \ } & \multicolumn{3}{c}{\ \hrulefill \ } \\                                       
I2013 & R.A. & Decl. 
 & $\mu_{X}$ & $\sigma \mu_{X}$ 
 & $\mu_{Y}$ & $\sigma \mu_{Y}$
 & V$_{Z}$ & $\Delta$V$_{Z}$  \\ \hline  
& & & & & & & & & 06/09/15 & 06/11/09 & 07/02/03 \\ \hline
  1   \ \dotfill \  &$    -0.13$&$    60.84$&$   1.37$&   0.09 &$   9.92$&   0.17
 &$ -49.15$&   2.42
  &        1.38 &        0.87 &    $\ldots$        \\                                                                                                                                          
  2   \ \dotfill \  &$     1.22$&$    60.82$&$   2.03$&   0.20 &$   8.38$&   0.17
 &$ -19.71$&   2.81
  &        0.13 &        0.09 &        0.63   \\                                                                                                                                          
  3   \ \dotfill \  &$     1.25$&$    60.76$&$   1.31$&   0.32 &$   9.94$&   0.55
 &$ -19.52$&   2.32
  &        0.12 &        0.15 &    $\ldots$        \\                                                                                                                                          
  4   \ \dotfill \  &$     1.23$&$    60.75$&$   2.10$&   0.73 &$   8.92$&   1.22
 &$ -17.49$&   1.37
  &        0.06 &        0.13 &    $\ldots$        \\                                                                                                                                          
  5   \ \dotfill \  &$    -0.60$&$    -2.68$&$   0.06$&   0.06 &$   0.48$&   0.13
 &$ 116.94$&   1.05
  &        0.09 &        0.15 &        0.19   \\                                                                                                                                          
  6   \ \dotfill \  &$     0.00$&$     0.01$&$   0.00$&   0.06 &$  -0.04$&   0.34
 &$ 117.85$&   1.19
  &        2.66 &        2.40 &        0.77   \\                                                                                                                                          
  7   \ \dotfill \  &$     0.00$&$    -1.06$&$  -0.40$&   0.42 &$   0.61$&   0.70
 &$ 117.89$&   0.94
  &        0.90 &        0.70 &    $\ldots$        \\                                                                                                                                          
  8   \ \dotfill \  &$    -0.10$&$    -0.83$&$   0.09$&   0.06 &$   1.11$&   0.26
 &$ 118.10$&   0.63
  &        1.54 &        2.08 &        1.25   \\                                                                                                                                          
  9   \ \dotfill \  &$    -0.25$&$     3.59$&$   0.04$&   0.04 &$   0.43$&   0.20
 &$ 118.37$&   1.90
  &        2.39 &        3.14 &        2.75   \\                                                                                                                                          
 10   \ \dotfill \  &$    -5.67$&$     6.27$&$  -0.27$&   0.19 &$  -0.57$&   0.51
 &$ 118.87$&   0.94
  &        0.43 &        0.91 &    $\ldots$        \\                                                                                                                                          
 11   \ \dotfill \  &$   -12.39$&$   -16.54$&$  -0.53$&   0.05 &$  -0.59$&   0.17
 &$ 121.26$&   0.52
  &        0.36 &        0.19 &    $\ldots$        \\                                                                                                                                          
 12   \ \dotfill \  &$   -21.85$&$   -16.50$&$  -1.09$&   0.12 &$  -0.24$&   0.19
 &$ 123.43$&   0.94
  &        0.15 &        0.13 &    $\ldots$        \\                                                                                                                                          
 13   \ \dotfill \  &$   -30.75$&$    16.98$&$  -1.53$&   0.07 &$   1.21$&   0.06
 &$ 130.56$&   0.98
  &        0.09 &        0.15 &        0.10   \\                                                                                                                                          
 14   \ \dotfill \  &$   -31.59$&$    15.26$&$  -1.47$&   0.05 &$   1.48$&   0.10
 &$ 132.14$&   1.12
  &        0.07 &        0.11 &        0.14   \\                                                                                                                                          
 15   \ \dotfill \  &$   -30.56$&$    19.15$&$  -1.44$&   0.13 &$   1.27$&   0.45
 &$ 133.06$&   0.52
  &    $\ldots$      &        0.03 &        0.02   \\
\hline
& & & & & & & & & 08/09/05 & 08/11/18 & 09/02/06 \\ \hline
 16   \ \dotfill \  &$     4.80$&$     8.32$&$  -0.22$&   0.35 &$  -0.50$&   0.82
 &$ 117.05$&   0.73
  &        0.54 &        0.81 &    $\ldots$        \\                                                               
  17   \ \dotfill \  &$    -0.71$&$    -0.27$&$  -2.02$&   1.01 &$  -2.09$&   0.78
 &$ 117.89$&   0.52
  &        2.29 &        1.18 &    $\ldots$        \\                                                               
  18   \ \dotfill \  &$    -0.69$&$    -0.40$&$  -0.03$&   0.55 &$  -0.08$&   0.96
 &$ 118.10$&   1.05
  &        2.75 &        2.33 &        1.50   \\                                                               
  19   \ \dotfill \  &$     0.00$&$     0.00$&$   0.00$&   0.41 &$   0.00$&   0.37
 &$ 118.96$&   1.48
  &        1.02 &        3.30 &        4.78   \\                                                               
  20  \ \dotfill \  &$    15.54$&$    27.45$&$   2.20$&   0.45 &$   1.15$&   0.63
 &$ 121.04$&   1.26
  &        0.72 &        0.62 &    $\ldots$        \\  \hline
\end{tabular}

\tablenotetext{1}{ Maser features detected in \iras1846. 
The features are designated as \iras1846:I2013-{\it N}, where {\it N} is the ordinal 
source number given in this column (I2013 stands for sources found by 
Imai et al. and listed in 2013).}
\tablenotetext{2}{ Relative value with respect to the location of 
the position-reference maser feature: \iras1846:I2013-{\it 6} and {\it 19} in the first three and the second three sessions, respectively.}
\tablenotetext{3}{ Relative value with respect to the motion of the position-reference 
maser feature: \iras1846:I2013-{\it 6} and {\it 19} in the first three and the second three sessions, respectively.}
\tablenotetext{4}{ LSR velocity.}
\end{table*}

%%%%% Table 5 %%%%%
\begin{deluxetable}{lrrrrr rrrrr}
\tablecaption{Parameters of the Detected  OH Maser Spots \label{tab:features_oh}}
\tabletypesize{\scriptsize}
\tablecolumns{7}
\tablehead{
& \multicolumn{5}{c}{On the Phase-referencing Based Image Cube}
&  \multicolumn{5}{c}{On Self-calibration Based Image Cube} \\
& \multicolumn{5}{c}{\ \hrulefill \ } & \multicolumn{5}{c}{\ \hrulefill \ } \\
\vlsr & \multicolumn{2}{c}{R.A.\ offset\tablenotemark{1} }& \multicolumn{2}{c}{Decl.\ offset\tablenotemark{1}}  
& $I_{\rm peak}$\tablenotemark{2} 
& \multicolumn{2}{c}{R.A.\ offset\tablenotemark{3} }& \multicolumn{2}{c}{Decl.\ offset\tablenotemark{3}}  
& $I_{\rm peak}$\tablenotemark{2} \\ 
(km s$^{-1}$) & \multicolumn{2}{c}{(mas)} &  \multicolumn{2}{c}{(mas)} & (Jy~beam$^{-1}$) 
& \multicolumn{2}{c}{(mas)} &  \multicolumn{2}{c}{(mas)} & (Jy~beam$^{-1}$)}
\startdata
\multicolumn{11}{c}{Red-shifted components} \\ \hline 
$   140.07$&$    1298.88$&   1.04 &$    -176.34$&   3.19 &         0.55  &$      -6.13$&   0.53 &$      -3.31$&   1.53 &         0.30  \\
$   139.89$&$    1296.94$&   1.11 &$    -177.57$&   3.30 &         0.78  &$      -7.55$&   0.45 &$      -3.12$&   1.19 &         0.47  \\
$   139.71$&$    1296.36$&   1.10 &$    -178.02$&   3.26 &         1.08  &$      -8.36$&   0.47 &$      -2.94$&   1.28 &         0.65  \\
$   139.53$&$    1296.71$&   0.98 &$    -176.43$&   2.86 &         1.31  &$      -8.45$&   0.31 &$      -2.94$&   0.82 &         0.81 \\
$   139.34$&$    1297.36$&   0.94 &$    -175.21$&   2.70 &         1.50  &$      -7.82$&   0.27 &$      -1.99$&   0.69 &         0.98 \\
$   139.16$&$    1298.20$&   0.83 &$    -175.40$&   2.31 &         1.53  &$      -7.49$&   0.14 &$      -1.62$&   0.33 &         1.09 \\
$   138.98$&$    1299.84$&   0.81 &$    -176.35$&   2.05 &         1.50  &$      -6.12$&   0.14 &$      -2.17$&   0.30 &         1.11 \\
$   138.80$&$    1303.61$&   0.89 &$    -177.21$&   2.03 &         1.43  &$      -2.41$&   0.24 &$      -2.16$&   0.47 &         1.01 \\
$   138.62$&$    1307.45$&   1.02 &$    -177.49$&   2.10 &         1.42  &$       1.95$&   0.29 &$      -1.86$&   0.55 &         1.00 \\
$   138.44$&$    1309.80$&   1.18 &$    -176.97$&   2.18 &         1.39  & \multicolumn{5}{c}{...} \\
$   138.26$&$    1309.68$&   1.18 &$    -178.61$&   1.98 &         1.32  & \multicolumn{5}{c}{...}  \\
$   138.07$&$    1311.11$&   1.23 &$    -181.22$&   2.09 &         1.29   & \multicolumn{5}{c}{...} \\
$   137.89$&$    1312.25$&   1.46 &$    -184.48$&   2.38 &         1.17  & \multicolumn{5}{c}{...}  \\
$   137.71$&$    1316.30$&   1.34 &$    -183.16$&   2.32 &         1.15  & \multicolumn{5}{c}{...}  \\
$   137.53$&$    1322.16$&   1.42 &$    -182.51$&   2.83 &         1.08  & \multicolumn{5}{c}{...}  \\
$   137.35$&$    1326.43$&   1.17 &$    -181.91$&   2.91 &         1.13  & \multicolumn{5}{c}{...}  \\
$   137.16$&$    1328.64$&   1.08 &$    -180.67$&   2.91 &         1.16  & \multicolumn{5}{c}{...}  \\
$   136.98$&$    1330.71$&   0.97 &$    -179.75$&   2.73 &         1.08  & \multicolumn{5}{c}{...}  \\
$   136.80$&$    1332.68$&   0.97 &$    -178.03$&   2.72 &         0.93  & \multicolumn{5}{c}{...}  \\
$   136.62$&$    1334.63$&   1.00 &$    -177.94$&   2.78 &         0.73  & \multicolumn{5}{c}{...}  \\
$   136.44$&$    1335.78$&   0.99 &$    -177.68$&   2.75 &         0.59  & \multicolumn{5}{c}{...}  \\
$   136.26$&$    1337.49$&   1.05 &$    -176.88$&   3.02 &         0.55  & \multicolumn{5}{c}{...}  \\
$   136.08$&$    1339.58$&   1.21 &$    -176.12$&   3.58 &         0.52  & \multicolumn{5}{c}{...}  \\
$   135.89$&$    1341.47$&   1.09 &$    -172.14$&   3.21 &         0.49  & \multicolumn{5}{c}{...}  \\
$   135.71$&$    1342.42$&   1.02 &$    -168.96$&   2.96 &         0.45  & \multicolumn{5}{c}{...}  \\
\hline \multicolumn{11}{c}{Blue-shifted components} \\  \hline
$   115.56$ & \multicolumn{5}{c}{...}  &$       1.29$&   0.43 &$     -25.10$&   1.17 &         0.39 \\
$   115.38$ & \multicolumn{5}{c}{...}  &$       0.69$&   0.26 &$     -29.94$&   0.69 &         0.58 \\
$   115.19$ & \multicolumn{5}{c}{...}  &$       0.94$&   0.27 &$     -28.32$&   0.71 &         0.75 \\
$   115.01$ & \multicolumn{5}{c}{...}  &$       1.24$&   0.14 &$     -27.34$&   0.36 &         0.78 \\
$   114.83$ & \multicolumn{5}{c}{...}  &$       2.21$&   0.15 &$     -22.51$&   0.35 &         0.67 \\
$   114.65$ & \multicolumn{5}{c}{...}  &$       2.05$&   0.19 &$     -17.93$&   0.46 &         0.56 \\
$   114.47$ & \multicolumn{5}{c}{...}  &$       1.55$&   0.30 &$     -13.95$&   0.81 &         0.69 \\
$   114.29$&$    1305.66$&   1.04 &$    -185.46$&   3.20 &         2.19 &$       0.39$&   0.37 &$     -14.45$&   1.03 &         1.09  \\
$   114.11$&$    1306.32$&   1.10 &$    -184.00$&   3.36 &         3.18  &$       0.34$&   0.35 &$     -13.84$&   0.99 &         1.83 \\
$   113.92$&$    1307.63$&   1.04 &$    -179.67$&   3.08 &         4.20  &$       1.73$&   0.26 &$      -9.90$&   0.72 &         2.68  \\
$   113.74$&$    1308.51$&   1.07 &$    -176.30$&   3.15 &         4.94  &$       2.95$&   0.19 &$      -5.41$&   0.51 &         3.33  \\
$   113.56$&$    1308.63$&   0.98 &$    -173.77$&   2.87 &         5.59 &$       3.45$&   0.18 &$      -2.39$&   0.49 &         3.81  \\
$   113.38$&$    1307.69$&   1.00 &$    -173.65$&   2.91 &         6.55 &$       2.47$&   0.18 &$      -2.08$&   0.47 &         4.35  \\
$   113.20$&$    1306.52$&   1.00 &$    -175.31$&   2.94 &         8.13  &$       1.17$&   0.21 &$      -3.42$&   0.57 &         5.06 \\
$   113.02$&$    1305.92$&   1.02 &$    -176.04$&   3.03 &        10.70 &$       0.49$&   0.28 &$      -4.20$&   0.77 &         6.63  \\
$   112.83$&$    1305.70$&   1.00 &$    -175.55$&   2.98 &        14.50  &$       0.30$&   0.34 &$      -4.28$&   0.93 &         9.01 \\
$   112.65$&$    1305.68$&   1.02 &$    -174.57$&   3.08 &        19.80  &$       0.28$&   0.38 &$      -3.54$&   1.05 &        12.30 \\
$   112.47$&$    1305.68$&   1.00 &$    -173.77$&   3.01 &        25.40  &$       0.33$&   0.35 &$      -2.35$&   0.97 &        16.60 \\
$   112.29$&$    1305.40$&   0.94 &$    -172.68$&   2.80 &        29.80  &$       0.06$&   0.28 &$      -1.48$&   0.75 &        20.80 \\
$   112.11$&$    1304.76$&   0.92 &$    -172.23$&   2.71 &        33.60  &$      -0.52$&   0.25 &$      -1.01$&   0.65 &        23.60 \\
$   111.93$&$    1304.31$&   0.96 &$    -173.01$&   2.84 &        35.50  &$      -1.08$&   0.26 &$      -0.93$&   0.68 &        26.10 \\
$   111.75$&$    1304.57$&   0.89 &$    -172.56$&   2.64 &        35.50  &$      -1.05$&   0.18 &$      -0.19$&   0.48 &        28.70 \\
$   111.56$&$    1305.25$&   0.95 &$    -171.96$&   2.82 &        32.40  &$      -0.37$&   0.15 &$       0.31$&   0.38 &        27.10 \\
$   111.38$&$    1305.83$&   0.89 &$    -172.08$&   2.58 &        22.50  &$       0.00$&   0.03 &$      -0.31$&   0.07 &        22.10 \\
$   111.20$&$    1305.36$&   1.05 &$    -168.04$&   2.65 &        12.20  &$       0.28$&   0.16 &$      -0.28$&   0.35 &        10.90 \\
$   111.02$&$    1303.76$&   2.03 &$    -165.21$&   4.37 &         3.66  &$       0.38$&   0.43 &$      -0.51$&   0.74 &         3.98 \\ 
\enddata

\tablenotetext{1}{ Position offset and its uncertainty with respect to the phase-tracking center of \iras1846.}
\tablenotetext{2}{ Peak intensity of the feature.}
\tablenotetext{3}{ Position offset and its uncertainty with respect to the origin of the image cube, which was 
obtained by the self-calibration procedures using the \vlsr$=$111.7\kms\ component as phase reference.}
\end{deluxetable}

%%%%% Table 6 %%%%%
\clearpage
\begin{table}[h]
\caption{Parameters of the Best Fit Three-dimensional Spatiokinematical Models of the 
\h2o Masers in the Central Slow Flow}
\label{tab:kinematic-model}

\begin{tabular}{lr@{}c@{$\pm$}lr} 
\hline \hline
Parameter & \multicolumn{2}{c}{Value} \\ \hline
$X_{\mbox{0}}$\tablenotemark{1} (mas) \dotfill \ & $-4.7$ & & 6.0 \\  
$Y_{\mbox{0}}$\tablenotemark{1} (mas) \dotfill \ & $-4.2$ & & 4.5 \\
$V_{\mbox{0x}}$\tablenotemark{1} (km~s$^{-1}$) \dotfill \ & $-$1.4 & & 2.1 \\  
$V_{\mbox{0y}}$\tablenotemark{1} (km~s$^{-1}$) \dotfill \ & 2.4 & & 2.7 \\
$V_{\mbox{0z}}$\tablenotemark{2} (km~s$^{-1}$) \dotfill \ 
& \multicolumn{3}{c}{124.7\tablenotemark{3}} \\
$D$ (kpc) \dotfill \ & \multicolumn{3}{c}{2.1\tablenotemark{4}} \\ \hline
rms residual $\sqrt{\chi^{2}}$ \dotfill \ & \multicolumn{3}{c}{1.23} \\ \hline 
\end{tabular}

\noindent
\tablenotetext{1}{ Relative value with respect to the position-reference maser feature. }
\tablenotetext{2}{ Velocity with respect to the local standard of rest.}
\tablenotetext{3}{ Assumed systemic radial velocity, which does not affect the $\chi^2$ value.}
\tablenotetext{4}{ Distance is completely covariant with the $z_{i}$ and $V_{\mbox{exp}}$($i$) and cannot be determined: $d\equiv$2.1 kpc.}
\end{table}

\clearpage
%%%%%%%%%%% Table 7 %%%%%%%%%%%%%%%%%%%
\begin{table*}
\caption{Results of Astrometry for the \h2o\ Maser Features Observed in 2008--2009}
\label{tab:secular-motion}
\scriptsize
\begin{tabular}{lr@{$\pm$}r r@{$\pm$}l r@{$\pm$}l r@{$\pm$}l r@{$\pm$}lrr} \hline \hline

Maser Feature\# & \multicolumn{2}{c}{$X_{\mbox \scriptsize 0}$\tablenotemark{1}} 
& \multicolumn{2}{c}{$Y_{\mbox \scriptsize 0}$\tablenotemark{1}}
& \multicolumn{2}{c}{$\mu_{X}$} & \multicolumn{2}{c}{$\mu_{Y}$} & \multicolumn{2}{c}{$\pi$} 
& \multicolumn{2}{c}{Deviation\tablenotemark{2}} \\
(\iras1846: I2013) & \multicolumn{2}{c}{[mas]} 
& \multicolumn{2}{c}{[mas]} & \multicolumn{2}{c}{[mas~yr$^{-1}$]}
& \multicolumn{2}{c}{[mas~yr$^{-1}$]} & \multicolumn{2}{c}{[mas]}  & $\sigma_{X}$ & $\sigma_{Y}$ \\ \hline
18\dotfill & 24.38 &    3.44 &   63.58 &    4.01 & $  -2.76$ &   0.48 & $  -7.35$ &  0.46 &  0.45 &  0.18 & 0.10 & 0.15 \\
19\dotfill & 32.99 &   10.49 &   56.96 &   3.69 & $  -3.73$ &   1.17 & $  -6.61$ &   0.41 &  0.58 &  0.29 & 0.09 & 0.21 \\
\hline
\end{tabular}
\tablenotetext{1}{ Position at the epoch J2000.0 with respect to the delay-tracking center (see main text).}
\tablenotetext{2}{ Mean deviation of the data points from the model-fit motion in unit of mas.}
\end{table*}

\clearpage
%%%%%%%%%%% Table 8 %%%%%%%%%%%%%%%%%%%
%\begin{table*}[h]
\begin{table}
\caption{Location and Three-dimensional Motion of \iras1846 in the Milky Way Estimated from the VLBA Astrometry}
\label{tab:MW-motion}

\scriptsize
\begin{tabular}{lr@{$\pm$}l@{}} \hline \hline
Parameter \hspace{3.8cm} & \multicolumn{2}{c}{Value} \\ \hline
Galactic coordinates, $(l,b)$ (deg)\tablenotemark{1} \dotfill  & \multicolumn{2}{c}{(31.01, $-$0.22)} \\
Heliocentric distance, $D$ (kpc)\tablenotemark{1} \dotfill  & 2.1 & 0.6 \\
Systemic LSR velocity, $V_{\rm sys}$ (km~s$^{-1}$)\tablenotemark{1} \dotfill & $124.7$ & 2.0 \\
Secular proper motion, $\mu_{\rm RA}$ (mas~yr$^{-1}$) \dotfill & $-3.24$ & 1.27 \\
\hspace*{30mm} $\mu_{\rm decl}$ (mas~yr$^{-1}$) \dotfill & $-6.47$ & 0.43 \\
$R_{\rm 0}$ (kpc)\tablenotemark{2} \dotfill  & \multicolumn{2}{c}{8.0} \\ 
$\Theta_{\rm 0}$ (km~s$^{-1}$)\tablenotemark{2} \dotfill & \multicolumn{2}{c}{240} \\
$(U_{\odot}, V_{\odot}, W_{\odot})$ (km~s$^{-1}$)\tablenotemark{3} \dotfill 
& \multicolumn{2}{c}{(7.5, 13.5, 6.8)} \\
$z_{\rm 0}$ (pc)\tablenotemark{4} \dotfill & \multicolumn{2}{c}{16} \\
Galactocentric distance, $R_{\rm gal}$ (kpc) \dotfill & 6.3 & 0.5 \\
$z$ (pc)\tablenotemark{5} \dotfill & 7 & 2 \\
$V_{R}$ (km~s$^{-1}$)\tablenotemark{6} \dotfill & 100 & 14 \\
$V_{\theta}$ (km~s$^{-1}$)\tablenotemark{6} \dotfill & 268 & 16 \\
$V_{z}$ (km~s$^{-1}$)\tablenotemark{6}  \dotfill & $-9$ & 11 \\ \hline
\end{tabular}

\tablenotetext{1}{ Input value for \iras1846.}
\tablenotetext{2}{ Input value for the Sun in the Milky Way.}
\tablenotetext{3}{ Motion of the Sun with respect to the local standard of rest, cited from \citet{fra09} (cf., \citealt{deh98}).}
\tablenotetext{4}{ Height of the Sun from the Galactic mid-plane, cited from \citet{ham95}.}
\tablenotetext{5}{ Height from the Galactic mid-plane.} 
\tablenotetext{6}{ Velocity vector components in cylindrical coordinates with respect to the Galactic Center, ($V_{R}, V_{\theta}, V_{z}$): in the radial direction, the azimuthal direction along the Galactic rotation, and the direction of the Galactic north pole, respectively.}
\end{table} 

\clearpage
%%%%% Figure 1 %%%%%

\clearpage
%%%%% Figure 2 %%%%%

\clearpage
%\end{figure}

\vspace{5mm}

%%%%% Figure 3 %%%%%

%%%%% Figure 4 %%%%%

%%%%% Figure 5 %%%%%

\end{document}